\documentstyle[12pt]{article}
             \newcommand{\om}{\omega}
\newcommand{\p}{\phi}                \newcommand{\rp}{r_+}
\newcommand{\cg}{\cal G}			 \newcommand{\cb}{\cal B}
\newcommand{\pr}{\partial}		     
			 
			 \newcommand{\G}{\Gamma}
\newcommand{\r}{\rho}                \newcommand{\vr}{\varrho}
\newcommand{\SS}{\Sigma}              \newcommand{\OO}{\Omega}
\newcommand{\tG}{{\tilde G}_{D + d}}
\newcommand{\tB}{{\tilde B}_{D + d}}

\newcommand{\rpl}{\big({\rp\over l}\big)}
\newcommand{\jl}{\big({J\over l}\big)}
\newcommand{\be}{\begin{equation}}
\newcommand{\ee}{\end{equation}}
\newcommand{\bq}{\begin{eqnarray}}
\newcommand{\eq}{\end{eqnarray}}
\newcommand{\rt}{1\over \sqrt{2}}
\newcommand{\trt}{1\over {2\sqrt{2}}}
\newcommand{\lrp}{l\over \rp}
\newcommand{\mpj}{\sqrt{M+\jl}}
\newcommand{\mnj}{\sqrt{M-\jl}}
\newcommand{\id}{{\rt}{\rpl}{{I_d}\over 2}}
\newcommand{\idi}{{\rt}\big({\lrp}\big){I_d}}

\begin{document}
\baselineskip= 24 truept
\begin{titlepage}
\title { $O(\tilde d, \tilde d)$ Transformations and \\
3D Black Hole}
\author{Abbas Ali and  Alok Kumar  \\
\\
 Institute of Physics, Bhubaneswar-751 005, India. \\
 email: abbas, kumar@iopb.ernet.in }
\date{}
\maketitle
\thispagestyle{empty}
\vskip .6in
\begin{abstract}
\vskip .2in
We generalize the results of a previous paper by one of the authors
to show a relationship among a class of string solutions through
$O(\tilde d, \tilde d)$ transformations. The results are applied
to a rotating black hole solution of
three dimensional general relativity discussed recently. We extend
the black hole
solution to string theory and show its connection with the three
dimensional black string with nonzero momentum through an
$O(\tilde d, \tilde d)$ transformation of the above type.
\end{abstract}
\bigskip
\flushright {IP/BBSR/92-80}
\flushright {December 1992}
\centerline{Submitted to {\it Mod. Phys. Lett. A}}
\vfil
\end{titlepage}
\eject

The search for the classical solutions of string theory and general
relativity in various space-time dimensions
has been an active area of research.
Recent developments in  understanding the
symmetries of the string effective actions \cite{mv}
has lead to considerable progress in this direction.
It has been shown that the string effective action is invariant
under an $O(\tilde d, \tilde d)$ group of
symmetry transformations \cite{mv},
when the background configuration is
independent of $\tilde d$ coordinates \cite{sen1}.
These symmetries have been utilized for generating
several new solutions of string effective action \cite{special}.

In this connection, an $O(\tilde d, \tilde d)$ transformation
was given in ref.\cite{alok} to show a
relationship between the background configuration of the ungauged
string actions to the gauged ones\cite{gr}.
In particular, it was shown that  such tansformations
can be used to go from the background fields
of the ungauged actions to the
ones which can be interpreted as black branes with zero linear
momentum along a number of directions. In this paper, we show that
the results of ref.\cite{alok} can be genralized further so that the
transformed solution corresponds to the branes with nonzero
momentum along these directions.

Our work was motivated partly by the
structure of  the rotating black hole solution of three
dimensional general relativity discussed recently in
ref.\cite{btz}(see also \cite{kogan}) which we extend to string
theory by inclusion of  antisymmetric tensor and
dilaton backgrounds. As an application of the results of the previous
paragraph, we show that the three dimensional rotating black hole
transforms to the moving black string \cite{hhs} by an $O(2, 2)$
transformation. We also find that the momentum of the black string is
proportional to the black hole angular momentum.

We start by reviewing some basic results.
It was shown in refs.\cite{mv} that the
string effective action and the equations of motion
are invariant under an $O(\tilde d, \tilde d)$
transformation which acts on the background fields as a linear
transformation $M \rightarrow {\Omega}M{\Omega ^T}$,
$\Phi \rightarrow \Phi$ where
\be
M = \left[\begin{array}{cc}
G^{-1}        &		 - G^{-1}B \\
BG^{-1}       & 	G - BG^{-1}B
\end{array} \right],\label{M}
\ee
and    $\Phi = \p - ln\sqrt{detG}$.
${\Omega}$ is an $O(\tilde d, \tilde d)$ matrix such
that ${\Omega ^T}\eta{\Omega} = \eta$, where
\be
\eta = \left[\begin{array}{cc}
0  &  I \\
I  &  0
\end{array} \right].\nonumber
\ee

Then in ref.\cite{alok}, the $O(\tilde d, \tilde d)$ transformation on a
general
background configuration in  $(D+d)$-dimensional space-time,
specified by the metric ($G_{D+d}^0$) and the antisymmetric
tensor ($B_{D+d}^0$), of the form\cite{gr}:
\be
G_{D+{d}}^0\; =
	\;\left (\matrix {{\G_s} & {{1\over 2}{\G}_1^T} &
	{{1\over 2}{\G}_2^T}\cr
{} & {} & {} \cr
 {{1\over 2}{\G}_1} & {I_d} & {\SS^T}\cr
{} & {} & {} \cr
{{1\over 2}{\G}_2} & {\SS} & {I_d}\cr }\right )\label{alokg},
\ee
and
\be
B_{D+{d}}^0\; =
	\;\left (\matrix {{\G_a} & {-{1\over 2}{\G}_1^T} &
	{{1\over 2}{\G}_2^T}\cr
{} & {} & {} \cr
 {{1\over 2}{\G}_1} & {0} & {\SS^T}\cr
{} & {} & {} \cr
 {-{1\over 2}{\G}_2} & {-\SS} & {0}\cr }\right )\label{alokb},
\ee
were studied. In eqns.(\ref{alokg}) and (\ref{alokb})
 $\G_{s,a} = \pm {1\over 2} {(\G \pm \G^T)}$,
are (anti-)symmetric parts of a $(D-d)\times(D-d)$ matrix $\G$.
$\G_{1, 2}, \SS$ are $d\times (D-d)$ and $d\times d$ matrices
respectively. The backgrounds in eqns.(\ref{alokg})-(\ref{alokb})
depend only on the first $(D-d)$ coordinates
$x^i \{i = 1, \cdots, (D-d)\}$. It was shown that the
above backgrounds transform by an $O(\tilde d, \tilde d)$
$(\tilde d = 2d)$ transformation of the form,
\be
{\OO}^0 = \left (\matrix { {I_{D-d}} & {0} & {0} & {0} &
	{0} & {0} \cr
{0} & {I_d\over 2} & {-{I_d\over 2}} & {0}
& {-{I_d\over 2}} & {-{I_d\over 2}}\cr
{0} & {I_d\over 2} & {{I_d\over 2}} & {0}
& {{I_d\over 2}} & {-{I_d\over 2}}\cr
{0} & {0} & {0} & {{I_{D-d}}}
& {0} & {0}\cr
{0} & {-{I_d\over 2}} & {-{I_d\over 2}} & {0}
& {{I_d\over 2}} & {-{I_d\over 2}}\cr
{0} & {I_d\over 2} & {-{I_d\over 2}} & {0}
& {{I_d\over 2}} & {{I_d\over 2}}\cr }\right )\label{alokom},
\ee
to the following metric and antisymmetric tensor fields,
\be
	{\tG}^0 =
	\left (\matrix {{G_D^0} &  {0}\cr
 	{0} & {I_d} \cr }\right )\label{alokg1},
\ee
\be
	{\tB}^0 =
	\left (\matrix {{B_D^0} &  {0}\cr
 	{0} & {0} \cr }\right )\label{alokb1},
\ee
where
\be
G_D^0 =
	\;\left (\matrix { {[\G_s-{1\over 4}
	{{\G_1}^T}{(I_d+\SS)}^{-1}{\G_2}} &
	{{1\over 2}[{{\G}_1^T}{(I_d+\SS)}^{-1}} \cr
	{-{1\over 4}{{\G_2}^T}{(I_d+\SS^T)}^{-1}{\G_1]}} &
	{-{{\G}_2^T}{(I_d+\SS^T)}^{-1}]} \cr
	{} & {} \cr
 	{{1\over 2}[{(I_d+\SS^T)}^{-1}{\G}_1} &
	{{1\over 2}{\left[{(I_d-\SS)}{(I_d+\SS)}^{-1} \right.}} \cr
	{-{(I_d+\SS)}^{-1}{{\G}_2]}} &
	{\left. +{(I_d+\SS^T)}^{-1}{(I_d-\SS^T)}\right]}	\cr}
 \right )\label{alokg2},
\ee
\be
B_D^0 =
	\left (\matrix {
	{[\G_a+{1\over 4}{{\G_1}^T}{(I_d+\SS)}^{-1}{\G_2}} &
	{-{1\over 2}{[{\G}_1^T}{(I_d+\SS)}^{-1}} \cr
	{-{1\over 4}{{\G_2}^T}{(I_d+\SS^T)}^{-1}{\G_1}]} &
	{+{{\G}_2^T}{(I_d+\SS^T)}^{-1}]} \cr
	{} & {} \cr
 	{{1\over 2}[{(I_d+\SS^T)}^{-1}{\G}_1} &
	{{1\over 2}{\left[-{(I_d-\SS)}{(I_d+\SS)}^{-1} \right.}} \cr
	{+{(I_d+\SS)}^{-1}{{\G}_2}]} &
	{\left. +{(I_d+\SS^T)}^{-1}{(I_d-\SS^T)}\right]}	\cr}
 \right ).\label{alokb2}
\ee
Corresponding dilaton field transformation is
${\tilde\p}_{D+d}^0 = \p_{D+d}^0 - ln[det(I_d+\SS)]\label{alokp1}$.

The background fields in eqns.(\ref{alokg})-(\ref{alokb})
 correspond to the ungauged
string action \cite{gr}
whereas the ones in eqn.(\ref{alokg1})-(\ref{alokb1})
 correspond to the gauged
ones with respect to the vector gauging \cite{gr}.
When the backgrounds in
eqn.(\ref{alokg})-(\ref{alokb})  represent the SL(2,R) WZW model,
then the backgrounds (\ref{alokg2})-(\ref{alokb2})  describe the
two dimensional black hole \cite{mandal,witten}. Moreover
the structure of the metric and antisymmetric
tensor in eqns.(\ref{alokg1})
and (\ref{alokb1}) point out a complete
decoupling of $d$ of the coordinates.
Hence the transformed background is a p-brane solution with
zero momentum along these directions. An interesting aspect of
the above $O(\tilde d, \tilde d)$ transformation
is that it does not depend on the
background configuration and is represented by a unique matrix in a
given space-time dimension.

We now generalize these results and obtain the
backgrounds with nonzero momentum of the p-brane. More precisely, we
obtain the $O(\tilde d, \tilde d)$ transformed background from the
following metric and antisymmetric tensor fields:
\be
G_{D+d} =
	\left(\matrix {{\G_s} & {{\trt}{\mnj}{\G}_1^T} &
	{{\trt}{\mpj}{\G}_2^T}\cr
{} & {} & {} \cr
 {{\trt}{\mnj}{\G}_1} & {{{1\over 2}(M - \jl)}I_d} &
 {{{1\over 2}{\rpl}^2}\SS^T}\cr
{} & {} & {} \cr
{{\trt}{\mpj}{\G}_2} &
{{{1\over 2}{\rpl}^2}\SS} &
{{{1\over 2}(M + \jl)}I_d}\cr }\right),\label{alokg3}
\ee
and
\be
B_{D+{d}}\; =
	\;\left(\matrix {{\G_a} & {-{\trt}{\mnj}{\G}_1^T} &
	{{\trt}{\mpj}{\G}_2^T}\cr
{} & {} & {} \cr
 {{\trt}{\mnj}{\G}_1} & {0} & {{{1\over 2}{{\rpl}^2}}\SS^T}\cr
{} & {} & {} \cr
{-{\trt}{\mpj}{\G}_2} &
{-{{1\over 2}{{\rpl}^2}}\SS} & {0}\cr }\right),\label{alokb3}
\ee
where $\rp^2 = Ml^2\sqrt{1 - \big({J\over Ml}\big)^2}$.
The $O(\tilde d, \tilde d)$ transformation is a generalization of the one
given in eqn.(\ref{alokom}) and is given by $\OO = $
\bq
\left(\matrix {
{I_{D-d}} & {0} & {0} & {0} &	{0} & {0} \cr
{0} & {\id} & {-\id} & {0}
& {-\idi} & {-\idi}\cr
{0} & {\id} & {\id} & {0}
& {\idi} & {-\idi}\cr
{0} & {0} & {0} & {{I_{D-d}}}
& {0} & {0}\cr
{0} & {-\id} & {-\id} & {0}
& {\idi} & {-\idi}\cr
{0} & {\id} & {-\id} & {0}
& {\idi} & {\idi}\cr }\right).\label{newom}
\eq

To obtain the transformed background, we
first notice that the $G_{D+d}^0$,
$B_{D+d}^0$ in eqns.(\ref{alokg}) and (\ref{alokb})
are related to  $G_{D+d}$ and $B_{D+d}$  in eqns.(\ref{alokg3})
and (\ref{alokb3}) by a constant coordinate transfomration
corresponding  to the $O(\tilde d, \tilde d)$ transformation:
\be
	A_c\; =
	\;\left (\matrix {{A} &  {0}\cr
 	{0} & {A^{-1 T}} \cr }\right ),
\ee
where
\be
	A  =
	\left (\matrix {{I_{D-d}} & {0} & {0}\cr
	{} & {} & {}\cr
	{0} & {\sqrt{2\over {M-\jl}}}I_d &  {0}\cr
    {} & {} & {}\cr
 	 {0} & {0} & {\sqrt{2\over {M+\jl}}}I_d \cr }\right ).
\ee
Then, using $\OO^0 M_0 {\OO^0}^T = \tilde{M_0}$ and demanding
$\OO M \OO^T = \tilde{M}$,
where $M_0, {\tilde M}_0, M,$ and $\tilde M$
are defined analogous to eqn.(\ref{M}),
one can prove that $\tilde{M_0}$ is related to $\tilde{M}$ by an
$O(\tilde d, \tilde d)$ transformation:
\be
	\tilde{A_c}\; =
	\;\left (\matrix {{\tilde{A}} &  {0}\cr
 	{0} & {\tilde{A}^{-1 T}} \cr }\right ),
\ee
where
\be
	\tilde{A} =
	\left (\matrix {{I_{D-d}} & {} & {0} & {} & {0}\cr
{} & {} & {} & {} & {}\cr
	{0} & {} & {{\sqrt{Ml^2+Jl}+\sqrt{Ml^2-Jl}}\over {2\rp}}I_d
	& {}  & {{\sqrt{Ml^2+Jl}-\sqrt{Ml^2-Jl}}\over {2\rp}}I_d\cr
{} & {} & {} & {} & {}\cr
    {0} & {} & {{\sqrt{Ml^2+Jl}-\sqrt{Ml^2-Jl}}\over {2\rp}}I_d
    & {} & {{\sqrt{Ml^2+Jl}+\sqrt{Ml^2-Jl}}\over {2\rp}}
	I_d\cr}\right).
	\label{coordtr}
\ee

The transformed fields $({\tG}$, ${\tB})$ can therefore
be obtained by a constant
coordinate transformation given in eq.(\ref{coordtr}).
For example, for the case when $\G = I$, $\G_1 = \G_2 = 0$
the new metric is given by	$\tG =$
\be
	\left(\matrix {{I} & {} & {0} & {0}\cr
{} & {} & {} & {}\cr
{0} & {} & {1\over 2}\left [{{(M({\lrp})^2I_d-\SS)}{(I_d+\SS)}^{-1}}
								\right.
& - {({\lrp})^2{\jl}}{1\over 2}\left[{(I_d+\SS)}^{-1}\right.    \cr
{} & {} & \left. +{(I_d+\SS^T)}^{-1}{(M({\lrp})^2I_d-\SS^T)}\right]
& \left.{ + (I_d + \SS^T)^{-1}}\right]\cr
{} & {} & {} & {}    \cr
{0} & {} & -{({\lrp})^2\jl}{1\over 2}\left[{(I_d+\SS)}^{-1}\right. &
{1\over 2}\left[{{(M({\lrp})^2I_d+\SS)}{(I_d+\SS)}^{-1}}\right. \cr
{} & {} & \left.+{(I_d+\SS^T)}^{-1}\right] &
\left.+{(I_d+\SS^T)}^{-1}{(M({\lrp})^2I_d+\SS^T)}\right] }\right)
\label{alokg4}.
\ee
General expression can also be obtained starting from
eqns.(\ref{alokg1})-(\ref{alokb1}).

It is now observed that the transformed metric has off-diagonal
terms in the d-directions which were earlier decoupled. As a result,
the new solution can be interpreted to represent a p-brane with
nonzero momentum along these directions. This interpretation is
further strengthened by the structure of the singularity for the case
when $J \neq 0$. Since the metric in eqns. (\ref{alokg1})
and (\ref{alokg4}) are related by a constant coordinate
transformation (\ref{coordtr}), the scalar curvature does not
depend on $J$. The inclusion of $J$, therefore, does not generate
any new singularity apart from the usual one at
$det(I + \SS) = 0$ for the case when $\p_{D+d}^0 = constant$.
However, it generates momentum along some of the isometry directions.

As stated earlier, the above generalization was motivated by the
space-time structure of the three dimensional rotating
black hole \cite{btz}.
This solution  has the metric of the form in eqn.(\ref{alokg3}).
To show this, we start by
writing down the three dimensional rotating black hole solution
of \cite{btz}. It is given by the metric:
\be
ds^2 = -N^2dt^2 + \r^2(N^\p dt + d\p)^2 +
\big({r\over \r}\big)^2N^{-2}dr^2
\label{btz}
\ee
where
\be
N^2 = \big({r\over \r}\big)^2\big({{r^2 - \rp^2}\over l^2}\big);
N^{\p} = - {J\over
{2\r^2}}; \r^2 = r^2 + {1\over 2}\big(Ml^2 - \rp^2\big).\nonumber
\ee
This metric satisfies the Einstein equations for the
comological constant $\lambda = -{1\over l^2}$.
The metric in eqn.(\ref{btz}) can be transformed
by a coordinate transformation,
$r = r_+ coshw$ and an orthogonal transformation
to
\be
\cg = \left(\begin{array}{cccc}
1 & {} & 0 & 0\\
0 & {} & {1\over 2}(M - \jl) & {1\over 2}(\rpl^2cosh 2\om) \\
0 & {} & {1\over 2}(\rpl^2cosh 2\om)  & {1\over 2}(M + \jl)
\end{array} \right)\label{btz1}
\ee
which we observe to be of the same form
as in eqn.(\ref{alokg3}), when $\G_1 =
\G_2 = 0$, $\G = 1$, and $\SS = cosh\om$.

We now extend this solution to string theory and obtain the
transformed solution which correspond
to the general result for ${\tG}$, ${\tB}$.
We find that it is precisely the
moving black string of ref.\cite{hhs}.
The  extension of the solution in eqn.(\ref{btz1}) is done by
first observing  that
the metric depends on a single space-time coordinate.
The string effective action with
backgrounds dependent on a single space-time coordinate were
studied in refs.\cite{mv}.
In such cases, the gauge symmetries of the string effective action
can be used to write the metric and the antisymmetric tensor in the
form
\be
{\cg} =  \left(\begin{array}{cc}
-1 & 0 \\
0  & G
\end{array} \right)
{\rm and}\;\;
{\cb} =  \left(\begin{array}{cc}
0  & 0 \\
0  & B
\end{array} \right)\label{ungauged00}
\ee
respectively.
Then the  equations of motion, following from the string effective
action are,
\be
(\dot{\Phi})^2 + {1\over 4}Tr[\pr _0G\pr _0G] + {1\over
4}Tr[G^{-1}(\pr _0B)G^{-1}(\pr _0B)] - \Lambda = 0, \label{mv1}
\ee
\be
(\dot{\Phi})^2 - 2\ddot{\Phi} -
{1\over 4}Tr[\pr _0G\pr _0G] -
{1\over4}Tr[G^{-1}(\pr _0B)G^{-1}(\pr _0B)]
- \Lambda - {{\pr \Lambda}\over
 {\pr \Phi}}= 0, \label{mv2}
\ee
\be
- \dot{\Phi}\dot{G} + G\pr _0(G^{-1}\dot{G}) - \dot{B}G^{-1}\dot{B}
= 0,\label{mv3}
\ee
\be
\dot{\Phi}\dot{B} - \ddot{B} + \dot{B}G^{-1}\dot{G} +
\dot{G}G^{-1}\dot{B} = 0.\label{mv4}
\ee

For our case, solution of the string
effective action is obtained by making
an ansatz that metric is the same as the black hole
metric of general relativity, i.e. eqn.(\ref{btz1})
and the dilaton $\p = c_0$ is a constant. The equations
of motion (\ref{mv1})-(\ref{mv4}) then reduce
to a set of differential equations for
the only independent component of the
antisymmetric tensor field. The
solution, upto a constant shift, is given by
\be
B = {1\over 2} \left[\begin{array}{cc}
0		& - \rpl^2cosh 2\om \\
\rpl^2cosh 2\om  & 0
\end{array} \right].\label{anti1}
\ee
The corresponding value of the
cosmological constant is $\Lambda = 4$.
Therefore, starting with a three dimensional rotating black hole
metric of general relativity,
we have obtained a consistent background
for the string effective action given by the
eqns.(\ref{btz1}) and (\ref{anti1})
together with a constant dilaton.
This solution corresponds to  a three dimensional
rotating axion black hole.

Now, the connection of the transformed backgrounds for this case to
the black string with nonzero momentum
 is established by writing down
the transformed fields. For the present  case we get,
\be
\tilde {\cg} = \left[\begin{array}{cc}
 -1 & 0 \\
0 & \tilde G
\end{array} \right]
\ee
where
\be
\tilde G = {1\over {1 + cosh 2\om }} \left[\begin{array}{cc}
M({\lrp})^2 - cosh 2\om & - ({\lrp})^2\jl \\
{} & {}\\
- {\lrp}^2\jl               & M{\lrp}^2 + cosh 2\om
\end{array} \right]\label{gauged1}
\ee
and $\tilde B = 0$.
The transformed dilaton background is given by
$\tilde {\p} = c_0 - ln cosh^2{\om}$.

To interpret the solution in  eqn.(\ref{gauged1}) we make the
coordinate transformation ${{\vr}\over m_0} = cosh^2{\om}$.
Then the dilaton background is given by,
$		\p = c_1 - ln{\vr}$
and the invariant distance can be written as:
\bq
ds^2 =& {{d\vr^2}\over {4\vr^2(1 - {m_0\over \vr})}} +
\left[{1 + m_0{{{1\over 2}(M({\lrp})^2 - 1)}\over
\vr}}\right]dx^2\nonumber\\
&- \left[{1 - m_0 {{{1\over 2}(M({\lrp})^2 + 1)}\over
\vr}}\right]dy^2
- {{m_0({\lrp})^2{\jl}}\over \vr}dxdy,\label{distance}
\eq
which matches precisely with the moving black
string of ref.\cite{hhs} for
$cosh^2{\alpha} = {1\over 2} ({{Ml^2\over r_+^2} + 1})$.

To conclude, we have presented a relationship among a class
of string backgrounds through
$O(\tilde d, \tilde d)$ transformations and
given an explicit example of three dimensional rotating
black hole. It will be interesting to apply our results to
various other examples. An interesting case may be the
investigation of four dimensional 2-branes,  discussed in
ref.\cite{2brane}, and its relationship with the four
dimensional rotating black hole. We hope to return to
this example in future.

\begin{flushleft}
{ACKNOWLEDGEMENT}
\end{flushleft}
We thank S.K. Kar  for discussions.

\vfil
\eject
\end{document}